\documentclass[conference]{IEEEtran}
\IEEEoverridecommandlockouts

\usepackage{cite}
\usepackage{amsmath,amssymb,amsfonts}
\usepackage{algorithmic}
\usepackage{graphicx}
\usepackage{textcomp}
\usepackage{xcolor}
\usepackage[hidelinks]{hyperref}
\usepackage{booktabs}
\def\BibTeX{{\rm B\kern-.05em{\sc i\kern-.025em b}\kern-.08em
    T\kern-.1667em\lower.7ex\hbox{E}\kern-.125emX}}

\title{Assessment of Latent Pedestrian--Vehicle Interaction Risk Profiles at Midblock Crossing in VR\\
\large\textbf{hEART 2026: 14th Symposium of the European Association for Research in Transportation, Paris, France}\\[-1.2em]
\thanks{This research is conducted under the project ``Virtual Reality to study the role sOcIal Conformity in the acceptance of Autonomous vehicles (VeROnICA),'' funded by the Economic and Social Research Council (United Kingdom).}
}

\author{%
\IEEEauthorblockN{%
Rulla Al-Haideri$^{1}$,
Bilal Farooq$^{1}$,
Elisabetta Cherchi$^{2}$%
}
\IEEEauthorblockA{%
$^{1}$\textit{Laboratory of Innovations in Transportation (LiTrans)}, \textit{Toronto Metropolitan University}, Canada\\
Email: rullaalhaideri@torontomu.ca \quad bilal.farooq@torontomu.ca%
}
\IEEEauthorblockA{%
$^{2}$\textit{New York University Abu Dhabi}, United Arab Emirates\\
Email: elisabetta.cherchi@nyu.edu%
}
}

\begin{document}
\maketitle
\begin{abstract}
Pedestrian safety at midblock crossings is a critical concern in mixed traffic environments where autonomous vehicles (AVs) and human-driven vehicles (HDVs) share the road. Pedestrians often infer intent from vehicle motion in AV encounters, making them vulnerable to small shifts in conflict margins. This study investigates whether virtual reality (VR) crossing sessions separate into distinct interaction risk profiles and whether AV-only sessions shift profile prevalence compared to HDV-only sessions.
Using large-scale immersive VR experiments from Toronto, Canada, and Newcastle, England, we compute surrogate safety measures (SSMs) and apply latent profile analysis (LPA) to identify distinct pedestrian crossing stances, ranging from risk-accepting to highly cautious.
Key findings show that Newcastle exhibits a higher prevalence of high-urgency risk profiles in AV-only sessions, indicating that AVs contribute to higher-risk encounters. In contrast, Toronto shows no significant difference between AV-only and HDV-only sessions, suggesting that contextual factors influence the impact of AVs on pedestrian safety.

\end{abstract}


\section{Introduction}
A traffic conflict is an observable interaction in which a collision would have happened if at least one road user had not taken an evasive action (\textit{e.g.}, braking, swerving, stopping, or accelerating) \cite{LaureshynVarhelyi2018}. These conflicts vary in severity, depending on the time and space margins between road users.
When these margins become small enough, crash avoidance may depend on rapid or substantial evasive manoeuvres. In this case, the interaction becomes safety-critical and could escalate to a crash if avoidance is no longer possible.
This severity is commonly operationalized using surrogate safety measures (SSMs). 
SSMs occur more frequently than crashes and preserve the kinematic structure of interactions. 
This makes them suitable for comparing safety-relevant behaviour under different conditions.
A key challenge in pedestrian--vehicle safety evaluation is heterogeneity. A small share of encounters can dominate the safety-relevant tail. Therefore collapsing SSMs to a single mean or median can mask distinct risk regimes including rare but consequential near misses.
This issue becomes more prominent in autonomous vehicles (AVs) settings. In such encounters, pedestrians often cannot rely on driver eye contact or informal gestures and must infer intent primarily from vehicle motion and any external human--machine interface (eHMI) cues. 
Immersive virtual reality (VR) proides a safe and a controlled way to study pedestrian--vehicle interactions while preserving naturalistic movement and decision-making which support SSM-based safety assessment.
Prior work employed latent constructs and segmentation to capture heterogeneity in pedestrian behaviour and safety outcomes \cite{HosseiniShoabjarehMamdoohiNordfjaern2021}. 
Recently, LPA has been used to stratify interaction severity when multiple SSMs are considered jointly \cite{MajiGhosh2025}. However, midblock pedestrian--vehicle interactions particularly in pedestrian--AV contexts are still evaluated using single summary statistic. This aggregation can mask whether AV-only versus human-driven vehicle (HDV)-only scenarios shift the distribution of interaction risk, especially in the upper tail where safety margins are smallest.

This paper employs LPA as an exploratory segmentation method to summarize multi-indicator SSM heterogeneity to a small number of interpretable risk profiles. The use of LPA is justified using two points. First, LPA moves the evaluation beyond single indicator averages by identifying distinct urgency regimes in the joint SSM space. Second, it enables direct comparisons of how often high urgency regimes occurred across vehicle types and locations. 
We apply this approach to a large scale VR midblock crossing experiments conducted in Toronto, Canada and Newcastle, England. Our unit of analysis is the VR experiment session with unique variable values (\textit{e.g.}, low vehicle flow, snowy weather, and a median in the middle), summarized using session level conflict indicators. For details on experiments, sessions, and data collection campaign, we refer the reader to Nazemi et al. \cite{Nazemi2025}. We estimated LPA separately by location using two temporal indicator specifications: Model~A used $\min(\mathrm{MTTC})$ and the corresponding CTTC at the same time step, while Model~B used $\min(\mathrm{T}_2)$ and the corresponding CTTC. We then compared profile prevalence between AV-only and HDV-only sessions within each location by focusing on the top highest risk end of the profile structure.
The main objective of this paper is to identify and compare latent interaction risk profiles in VR midblock pedestrian--vehicle encounters using SSMs, and to examine how profile prevalence and structure vary by vehicle type and location. This paper addresses the following research questions:
\begin{enumerate}
    \item Within each location, what latent interaction risk profiles emerge when each VR session is represented by multiple temporal SSMs?
    \item Within each location, does the proportion of sessions assigned to the highest risk profile differ between AV-only and HDV-only scenarios?
    \item Across locations, are the latent profiles comparable in number and indicator patterns, and do AV-only versus HDV-only differences change when the indicator set is used?
\end{enumerate}
The remainder of the paper is organized as follows. Section~\ref{sec:methods}
presents the methods, including the data analysis and modelling framework.
Section~\ref{sec:results} reports the results and discussion. Section~\ref{sec:conclusions} concludes and outlines future work.

\section{Methods}
\label{sec:methods}
\subsection{Data Analysis}
This study uses two VR datasets collected in Toronto (109 participants) and Newcastle (249 participants). Each dataset includes time-stamped trajectories for pedestrians and simulated vehicle agents during a midblock crossing task on a two-lane road. Participants complete repeated sessions and each lasts for 60~s under systematically varied conditions. Within the two locations, each participant completed 12 to 24 sessions, with conditions varying by time of day, weather, vehicle type (AV versus HDV), presence of an eHMI for the AV, presence of a median, behaviour of avatar pedestrians, and traffic flow. The scenario structure is comparable across locations to support within location modelling under consistent task demands. 
Raw trajectories are recorded at 0.1~s and are post-processed and smoothed to reduce measurement noise, and then resampled to 1~s to create a uniform time series.
Three predictive temporal conflict indicators are calculated to quantify session-level interaction criticality under both collision course and non-collision course encounters. These include the modified time-to-collision (MTTC), time-to-arrival ($\mathrm{T}_2$), and closing time-to-collision (CTTC). 
For LPA, each 60~s session is represented by its most critical time step. This is defined as the minimum of a primary temporal indicator and the corresponding CTTC at that same time step. Two indicator sets are analyzed within each location. Model~A uses $\min(\mathrm{MTTC})$ with CTTC evaluated at the time of $\min(\mathrm{MTTC})$. Model~B uses $\min(\mathrm{T}_2)$ with CTTC evaluated at the time of $\min(\mathrm{T}_2)$.
MTTC is the time for two road users to collide under constant acceleration extrapolation, evaluated only when the pair is on a collision course~\cite{Ozbay2008}. In this paper, road users are represented as rectangles and collision times are computed using the corner--side procedure of Laureshyn et al.~\cite{Laureshyn2010}.
$\mathrm{T}_2$ is the predicted arrival time of the later road user to the avoided conflict point ~\cite{Laureshyn2010}. 
CTTC is the instantaneous closing time implied by the current separation distance divided by the line-of-sight closing speed (relative velocity projected onto the line of sight) under constant-velocity extrapolation~\cite{alhaideri2025cttc}. 
The estimation samples used in this study represent subsets of the full recorded datasets. 
Sessions are included in the LPA estimation sample only when both indicators required by the selected model are defined (finite after pre-processing). Prior to LPA, session level conflict indicators are log transformed to reduce right skew and then standardized using z-scores computed within each location and indicator model as follows:
\begin{equation}
z = \frac{\log(x)-\mu_{\text{loc,model}}}{\sigma_{\text{loc,model}}},
\end{equation}
where $\mu_{\text{loc,model}}$ and $\sigma_{\text{loc,model}}$ are the mean and standard deviation of $\log(x)$  computed from the sessions included in that model. The unit of analysis in this paper is a session, defined as a 60~s pedestrian crossing trial. It is summarized by session level minima of temporal surrogate indicators.

\subsection{Modelling Framework}
The VR procedure and control variables are coded consistently across Toronto and Newcastle. However, we estimate LPA separately by location because pooling requires assuming that the indicator distributions defining each profile are the same across sites. Testing and imposing cross site measurement invariance within profiles (and estimating a pooled multi-group LPA) is beyond the scope of this paper and is left for future work.
In each location, models with $K=1,2,\ldots,K_{\max}$ are evaluated, and solutions are discarded if any profile contained fewer than 5\% of sessions.
Classification quality is assessed using posterior membership probabilities. The distribution of maximum posterior probabilities is summarized, and average posterior probabilities of assignment of around 0.70 or higher are treated as indicative of acceptable classification~\cite{NaginOdgers2010}.
Among candidate solutions, the smallest number of profiles \(K\) that still provides an adequate fit is selected. First, the model with the lowest BIC is identified. The smallest \(K\) whose BIC is within 10 points of that best (lowest) BIC is then selected. The \(K\) is increased only if moving from \(K-1\) to \(K\) improves BIC by at least 2 points. Otherwise the more parsimonious \(K-1\) solution is kept.
After selecting \(K\), profile prevalence is reported separately for AV-only and HDV-only sessions for each location. Profiles are ranked only for consistent labelling (rank~1 as highest risk). These ranks are not used in model estimation or model selection. To rank profiles, a composite severity score is defined for each session as:
\begin{equation}
r_i = -\sum_{j \in \mathcal{J}} z_{ij},
\end{equation}
where \(\mathcal{J}\) is the set of standardized indicators used in the selected model.
For each session, the LPA model returns posterior membership probabilities for all \(K\) profiles. Each session is assigned to the profile with the largest posterior probability. Using these assignments, profile prevalence is computed separately for AV-only and HDV-only sessions per location. 
For each session $i$ in a given location and indicator model, the 2D feature vector is defined as:
\[
\mathbf{x}_i =
\begin{bmatrix}
z_{\text{loc,model}}\!\left(\log \, t_i\right) \\
z_{\text{loc,model}}\!\left(\log \, c_i\right)
\end{bmatrix}
\in \mathbb{R}^{2},
\]
where $t_i$ is the session level conflict indicator ($t_i=\min(\mathrm{MTTC})$ for Model~A or $t_i=\min(\mathrm{T}_2)$ for Model~B), and $c_i$ is the corresponding $\mathrm{CTTC}$ value evaluated at the time step where $t_i$ occurred. The $z_{\text{loc,model}}(\cdot)$ denotes z-score standardization computed within each location and indicator model using the sessions included in that model.
LPA represents the distribution of the session level indicator vector $\mathbf{x}_i$ as a mixture of $K$ latent profiles. Each session $i$ belongs to an unobserved profile $Z_i \in \{1,\ldots,K\}$ with mixing proportions $\pi_k=\Pr(Z_i=k)$, where $\sum_{k=1}^{K}\pi_k=1$. The mixture density is expressed by:
\begin{equation}
f(\mathbf{x}_i) = \sum_{k=1}^{K} \pi_k \, \phi\!\left(\mathbf{x}_i \mid \boldsymbol{\mu}_k, \boldsymbol{\Sigma}_k\right),
\end{equation}
where $\phi(\cdot \mid \boldsymbol{\mu}_k, \boldsymbol{\Sigma}_k)$ is the bivariate normal density for profile $k$ with mean $\boldsymbol{\mu}_k$ and covariance $\boldsymbol{\Sigma}_k$.
Model parameters $\{\pi_k,\boldsymbol{\mu}_k,\boldsymbol{\Sigma}_k\}_{k=1}^{K}$ are estimated by maximum likelihood using an expectation--maximization algorithm. For each session, posterior membership probabilities are obtained by:
\begin{equation}
\gamma_{ik}=\Pr(Z_i=k\mid \mathbf{x}_i)=
\frac{\pi_k\,\phi(\mathbf{x}_i\mid \boldsymbol{\mu}_k,\boldsymbol{\Sigma}_k)}
{\sum_{h=1}^{K}\pi_h\,\phi(\mathbf{x}_i\mid \boldsymbol{\mu}_h,\boldsymbol{\Sigma}_h)} ,
\end{equation}
and the session was assigned to the profile with the largest $\gamma_{ik}$.
All models are estimated in \textsf{R} using \texttt{tidyLPA}~\cite{vanLissa2018tidyLPA}.

\section{Results and Discussion}
\label{sec:results}
Table~\ref{tab:model_fit_by_location} summarizes the selected number of profiles, estimation sample sizes, and classification diagnostics for each model. 
The selected solutions ranged from $K=4$ (Newcastle Model~A) to $K=6$ (Newcastle Model~B and Toronto Model~A). The entropy summarizes how clearly sessions are assigned to a single latent profile, and values closer to 1 indicate clearer separation. The mean maximum posterior probability is the average across sessions of the largest profile membership probability, where values closer to 1 imply higher assignment certainty. As can be seen from the table, classification separation is strong for Newcastle Model~A (entropy $=0.9551$), Newcastle Model~B (entropy $=0.9465$), and Toronto Model~A (entropy $=0.9539$), with mean maximum posterior probabilities above 0.96. Toronto Model~B shows weaker separation (entropy $=0.8641$; mean maximum posterior probability $=0.9128$). This suggests that the \(\mathrm{T}_2\)-based indicator space in Toronto captures either a more continuous spectrum of risk or higher within-profile variability. 
As a result, profiles overlap more, and session-to-profile assignments are less stable.

\begin{table*}[t]
\centering
\caption{LPA model fit summary by location.}
\label{tab:model_fit_by_location}
\renewcommand{\arraystretch}{1.15}
\footnotesize
\begin{tabular}{@{}lllllllll@{}}
\toprule
Location & Model & $K$ & $N_{\mathrm{AV}}$ & $N_{\mathrm{HDV}}$ &
BIC & Entropy & Min.\ class prop.\ & Mean max post.\ \\
\midrule
NC & A & 4 & 445  & 242 & 616.84  & 0.9551 & 0.1645 & 0.9830 \\
NC & B & 6 & 1141 & 569 & 6024.75 & 0.9465 & 0.0696 & 0.9655 \\
TO & A & 6 & 607  & 584 & 2403.77 & 0.9539 & 0.0705 & 0.9677 \\
TO & B & 5 & 445  & 458 & 3717.29 & 0.8641 & 0.0554 & 0.9128 \\
\bottomrule
\end{tabular}\par
\vspace{2pt}
{\footnotesize\centering
\textit{Notes:} TO: Toronto, NC: Newcastle. $K$ is selected number of latent profiles. $N_{\mathrm{AV}}$ and $N_{\mathrm{HDV}}$ are numbers of sessions used for estimation in AV-only and HDV-only.\par}
\vspace{-7pt}
\end{table*}

\subsection{AV-only versus HDV-only differences in high criticality profile prevalence}
Table~\ref{tab:key_contrasts} presents contrasts in profile prevalence between AV-only and HDV-only sessions for (i) the highest interaction risk profile and (ii) the combined set of the top two highest interaction risk profiles. These contrasts focus on the upper tail of interaction risk, where smaller temporal margins are most safety-relevant.
As a general observation, Table~\ref{tab:model_fit_by_location} indicates strong profile separation in Newcastle and in Toronto under Model~A. 
In contrast, Toronto under Model~B shows weaker separation, with lower entropy and a lower mean maximum posterior probability, consistent with greater overlap in the \(\mathrm{T}_2\)-based indicator space.
In Newcastle, AV-only sessions exhibit a higher prevalence of the combined top two highest risk profiles under both indicator models. Under Model~A, the top two prevalence is higher in AV-only sessions (\(\Delta p=0.1313\), 95\% CI \([0.0245,\,0.2332]\)). Whereas the contrast for the single highest risk profile is not distinguishable (\(\Delta p=-0.0395\), 95\% CI \([-0.1239,\,0.0411]\)). Under Model~B, AV-only sessions show higher prevalence for both the highest risk profile (\(\Delta p=0.0809\), 95\% CI \([0.0182,\,0.1412]\)) and the combined top two profiles (\(\Delta p=0.1456\), 95\% CI \([0.0788,\,0.2100]\)). These findings suggest that AV-only conditions are associated with a larger share of sessions in high risk interaction regimes in Newcastle, implying smaller temporal safety margins for a non-trivial subset of encounters.
This result aligns with previous research. Chen et al.~\cite{Chen2020} found that pedestrians tend to accept larger gaps when interacting with AVs, potentially leading to more risky encounters if pedestrians misjudge the safety of the interaction. On the other hand, Velasco et al.~\cite{Velasco2021} demonstrated that AVs sometimes induce more cautious behaviour in pedestrians, but this effect is context dependent. In some cases, pedestrians accepted smaller gaps when interacting with HDVs, highlighting that pedestrians' decision making in crossing situations varies depending on the vehicle type and the surrounding context. Our results indicate that, in Newcastle, AVs are perceived as riskier in terms of interaction criticality. This suggests that pedestrians may misinterpret AV intent or feel more uncertainty during AV encounters compared to HDVs.
\begin{table*}[t]
\centering
\caption{Key contrasts in interaction risk profile prevalence between AV-only and HDV-only scenarios by location.}
\label{tab:key_contrasts}
\renewcommand{\arraystretch}{1.15}
\footnotesize
\begin{tabular}{@{}llp{3.6cm}lllllllll@{}}
\toprule
Location & Model & Contrast &
$k_{\mathrm{AV}}$ & $N_{\mathrm{AV}}$ & $p_{\mathrm{AV}}$ &
$k_{\mathrm{HDV}}$ & $N_{\mathrm{HDV}}$ & $p_{\mathrm{HDV}}$ &
$\Delta p$ & 95\% CI \\
\midrule
NC & A & Top-1 risk profile &
67 & 445 & 0.1506 &
46 & 242 & 0.1901 &
-0.0395 & [$-0.1239$, 0.0411]$^{*}$ \\
NC & A & Top-2 risk profiles (rank 1--2) &
200 & 445 & 0.4494 &
77 & 242 & 0.3182 &
0.1313 & [0.0245, 0.2332] \\
NC & B & Top-1 risk profile &
357 & 1141 & 0.3129 &
132 & 569 & 0.2320 &
0.0809 & [0.0182, 0.1412] \\
NC & B & Top-2 risk profiles (rank 1--2) &
489 & 1141 & 0.4286 &
161 & 569 & 0.2830 &
0.1456 & [0.0788, 0.2100] \\
TO & A & Top-1 risk profile &
126 & 607 & 0.2076 &
119 & 584 & 0.2038 &
0.0038 & [$-0.0611$, 0.0685]$^{*}$ \\
TO & A & Top-2 risk profiles (rank 1--2) &
175 & 607 & 0.2883 &
154 & 584 & 0.2637 &
0.0246 & [$-0.0472$, 0.0960]$^{*}$ \\
TO & B & Top-1 risk profile &
29 & 445 & 0.0652 &
21 & 458 & 0.0459 &
0.0193 & [$-0.0233$, 0.0618]$^{*}$ \\
TO & B & Top-2 risk profiles (rank 1--2) &
103 & 445 & 0.2315 &
100 & 458 & 0.2183 &
0.0131 & [$-0.0638$, 0.0899]$^{*}$ \\
\bottomrule
\end{tabular}

\vspace{2pt} 
\footnotesize\centering
\textit{Notes:} $k_{\mathrm{AV}}$ and $k_{\mathrm{HDV}}$ are numbers of sessions assigned to contrasted profile set (top-1 or top-2) in AV-only and HDV-only; $N_{\mathrm{AV}}$ and $N_{\mathrm{HDV}}$ are total numbers of sessions scenarios. $p_{\mathrm{AV}}=k_{\mathrm{AV}}/N_{\mathrm{AV}}$ and $p_{\mathrm{HDV}}=k_{\mathrm{HDV}}/N_{\mathrm{HDV}}$. $\Delta p = p_{\mathrm{AV}}-p_{\mathrm{HDV}}$ is reported with Newcombe--Wilson 95\% CIs. $^{*}$CI includes 0 (contrast is not distinguishable at the 95\% level).
\vspace{-7pt}
\end{table*}
In Toronto, AV-only versus HDV-only contrasts are small and not distinguishable for both indicator models. For Model~A, $\Delta p=0.0038$ for the Top-1 risk profile (95\% CI $[-0.0611,\,0.0685]$) and $\Delta p=0.0246$ for the top two profiles (95\% CI $[-0.0472,\,0.0960]$). For Model~B, $\Delta p=0.0193$ for the highest risk profile (95\% CI $[-0.0233,\,0.0618]$) and $\Delta p=0.0131$ for the top two profiles (95\% CI $[-0.0638,\,0.0899]$). Within the estimation samples used for each model, Toronto exhibits broadly similar prevalence of high-risk profiles across AV-only and HDV-only scenarios. However, non-distinguishable contrasts do not imply equivalence. They indicate that the available data do not rule out small differences in either direction.
Differences between Model~A and Model~B are consistent with their indicator definitions. Model~A relies on MTTC, which is defined only for collision-course events under the assumed motion model and therefore emphasizes imminent collision configurations. Model~B relies on $\mathrm{T}_2$ and CTTC, which remain informative in near-miss encounters because they characterize arrival timing and closing urgency even when a collision is not predicted. This distinction is consistent with the stronger AV-only versus HDV-only differences in Newcastle under Model~B and the weaker separation observed for Toronto Model~B in Table~\ref{tab:model_fit_by_location}.

\subsection{Risk profile prevalence and indicator structure}
\begin{figure}[t]
  \includegraphics[width=\columnwidth]{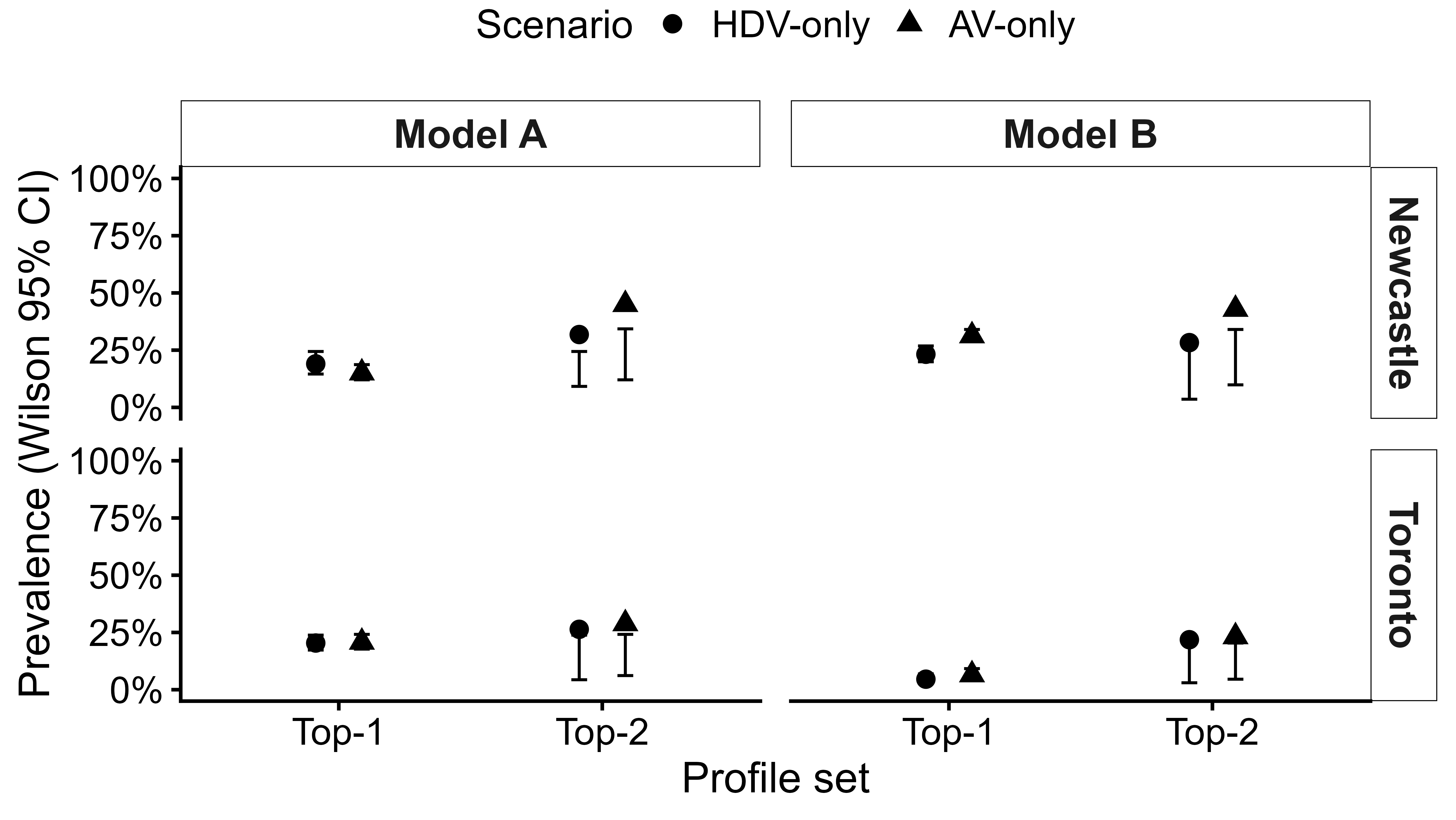}
  \vspace{-20pt} 
  \caption{Prevalence of the highest risk end of the latent profile structure by scenario type.}
  \label{fig:prev_toprisk}
\end{figure}

\begin{figure}[t]
  \includegraphics[width=\columnwidth]{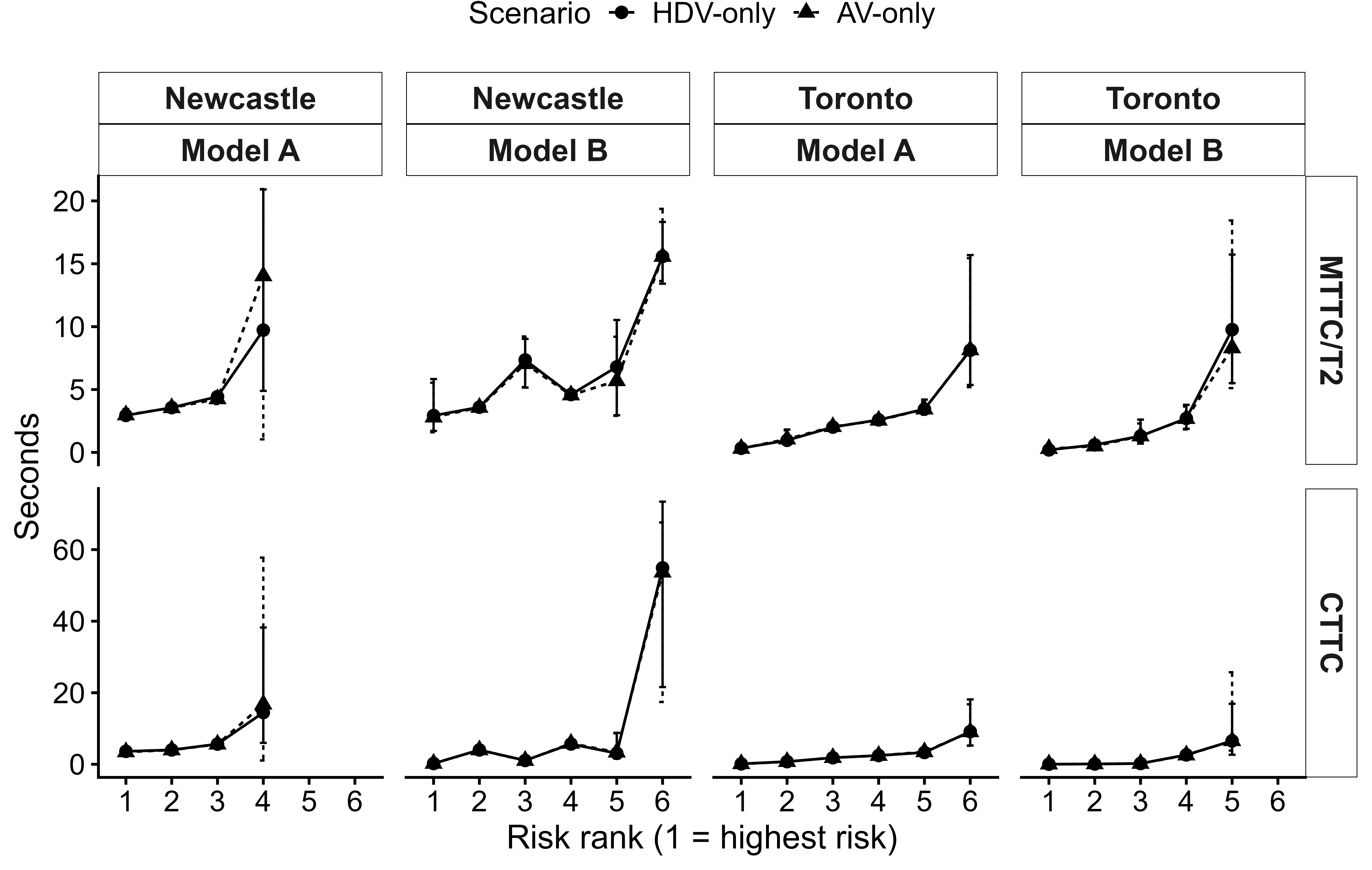}
  \vspace{-20pt} 
  \caption{Indicator levels by risk rank showing medians with 10th--90th percentile ranges.}
  \label{fig:cues_by_riskrank}
\end{figure}

Figure~\ref{fig:prev_toprisk} reports the prevalence of the highest risk end of the latent profile structure by scenario type (AV-only versus HDV-only). Prevalence is reported for the single highest risk profile (Top-1) and for the combined set of the two highest risk profiles (Top-2), with Wilson 95\% CIs.
In Newcastle, AV-only sessions exhibit a higher prevalence of high-risk interactions under Model~B. Both the Top-1 and Top-2 sets are more prevalent in AV-only sessions than in HDV-only sessions. This implies a shift toward smaller temporal safety margins when encounters involve AVs only. In Model~A, the contrast is lower and is more evident when the two highest risk profiles are considered jointly. This pattern is consistent with stronger separation when near-miss informative timing cues ($\mathrm{T}_2$ with CTTC) define the latent space.
In Toronto, prevalence estimates for AV-only and HDV-only sessions are similar for both indicator models. CIs overlap for both Top-1 and Top-2 sets. This confirms that the data do not support distinguishable differences in the prevalence of high-risk interactions by vehicle type within this location. The prevalence results suggest that AV-only versus HDV-only shifts in interaction risk are location dependent and are more detectable under indicator definitions that retain near-miss timing structure rather than emphasizing collision-course imminence alone.
Figure~\ref{fig:cues_by_riskrank} summarizes indicator levels within each risk rank and provides an internal consistency check for the latent ordering. Lower rank numbers (rank~1) correspond to smaller MTTC/$\mathrm{T}_2$ values and smaller CTTC values, implying shorter time margins and greater closing urgency. The monotonic ordering of indicator values across risk ranks provides a construct validity check for the composite severity ranking. This indicates that the profiles are aligned with the surrogate indicators used to define them rather than reflecting arbitrary label permutations.
In Newcastle, separation between adjacent risk ranks is visually clearer, particularly for Model~B. This aligns with the strong classification diagnostics discussed earlier. In Toronto under Model~B, indicator ranges overlap more across ranks, especially in the mid-risk range. This overlap is consistent with the lower entropy observed for this specification and suggests that the $\mathrm{T}_2$-based indicator space captures a more continuous spectrum of urgency in this context. To aid interpretation, we next describe what these ranks could represent as hypothesized pedestrian crossing stances expressed within a session.
\subsection{Interpretation of risk ranks and hypothesized pedestrian crossing stances}
Risk ranks represent clusters of instances during a session in temporal margin and closing urgency space. Each session is represented by the indicator values at its most critical time instance, defined by the minimum MTTC (Model~A) or minimum $\mathrm{T}_2$ (Model~B) and the corresponding CTTC at the same time step. We hypothesize that these ranks correspond to pedestrian crossing stances expressed within a session, ranging from risk-accepting (small temporal margins and high closing urgency) to highly cautious (large margins and weak closing). The plotted values summarize session minima. Lower ranks (rank~1) indicate sessions whose most critical moment has smaller temporal margins and higher closing urgency. Higher ranks indicate sessions whose minimum indicator values remain comparatively large.
\paragraph{Newcastle, Model~A (MTTC and CTTC; $K=4$)}
In Newcastle Model~A, ranks~1--3 are close in indicator magnitude. Rank~4 is distinct with much larger MTTC and CTTC.
Rank~1 (\emph{risk taker / urgent crosser}) shows MTTC and CTTC values around 3~s. This is consistent with when the pedestrian accepts a tight timing situation or commits late which require rapid resolution under urgent closing.
Rank~2 (\emph{assertive but controlled}) shows a slightly higher MTTC with a similar CTTC. This is consistent with a modestly larger collision course time margin while closing urgency remains comparable.
Rank~3 (\emph{typical / moderately cautious}) shows moderately higher MTTC and marginally higher CTTC. This is consistent with pedestrians where the minimum occurs with somewhat larger buffers and more recoverable interaction conditions.
Rank~4 (\emph{cautious / risk averse}) shows a sharp increase in MTTC and CTTC. This indicates sessions whose minimum collision course moment remains well-buffered. In this lowest risk regime, MTTC and CTTC are slightly higher in AV-only than in HDV-only sessions. This is consistent with larger temporal buffers (or weaker closing) when sessions fall into this configuration.

\paragraph{Newcastle, Model~B ($\mathrm{T}_2$ and CTTC; $K=6$)}
Newcastle Model~B shows greater variability between ranks because $\mathrm{T}_2$ remains defined in near miss encounters. This richer definition separates regimes where avoided conflict timing is tight from regimes where instantaneous closing urgency is high which allows these dimensions to decouple.
Rank~1 (\emph{risk taker / high urgency}) shows $\mathrm{T}_2$ values around 2~s with CTTC near 1~s. This is consistent with sessions where avoided conflict timing becomes tight under very urgent closing.
Rank~2 (\emph{assertive but anticipatory}) has a slightly higher $\mathrm{T}_2$ and distinctly higher CTTC than rank~1. This is consistent with tight timing but moderated closing urgency. For example, when the pedestrian commits under small margins but interaction dynamics already soften through speed adaptation or larger separation at the minimum-$\mathrm{T}_2$ moment.
Rank~3 (\emph{opportunistic / confident}) exhibits a much larger $\mathrm{T}_2$ while CTTC returns closer to rank~1. This is consistent with sessions where the pedestrian experiences urgent closing in distance (small CTTC) but arrival timing at the conflict point remains well separated (large $\mathrm{T}_2$). This could occur due to interaction geometry or because one road user passes the conflict point substantially earlier than the other.
Rank~4 (\emph{negotiator}) shows lower $\mathrm{T}_2$ than rank~3 with slightly higher CTTC. This is in line with sessions where timing becomes tighter but closing urgency is partially moderated through interaction adjustment.
Rank~5 (\emph{AV-trusting or AV-sensitive}) shows scenario dependent separation. In the HDV-only sessions it exhibits larger $\mathrm{T}_2$ than AV-only sessions, while CTTC remains similar (around 2~s). This implies that in these sessions the pedestrians operate with smaller avoided conflict timing margins in AV-only encounters at comparable closing urgency. This is consistent with either greater willingness to accept tighter timing against AVs or different AV approach and yielding dynamics that compress timing without increasing immediate closing urgency.
Rank~6 (\emph{very cautious / highly conservative}) is distinct with very large $\mathrm{T}_2$ (around 15~s) and very large CTTC (around 60~s). This is consistent with sessions where even the minimum remains highly buffered, reflecting very conservative waiting for large gaps or low exposure to close vehicle approaches.

\paragraph{Toronto, Model~A (MTTC and CTTC; $K=6$)}
In Toronto Model~A, MTTC increases steadily from ranks~1 through~5. CTTC values are comparatively similar and increase only marginally. This indicates that ranks primarily separate sessions by how small the collision course time margin becomes at the most critical moment.
Rank~1 (\emph{risk taker / urgent crosser}) reflects the smallest MTTC and low CTTC. This is in line with tight collision course timing under urgent closing.
Ranks~2--3 (\emph{assertive to typical}) reflect progressively larger MTTC with similar closing urgency. This is consistent with more recoverable but still interaction relevant sessions.
Ranks~4--5 (\emph{cautious}) reflect larger MTTC and slightly higher CTTC. This reflects sessions where the minimum occurs with more buffered timing.
Rank~6 (\emph{very cautious / risk averse}) is distinct with larger MTTC (around 10~s) and higher CTTC (around the mid-teens). This indicates sessions whose minimum collision course moment remains well buffered.

\paragraph{Toronto, Model~B ($\mathrm{T}_2$ and CTTC; $K=5$)}
Toronto Model~B shows a similar trend toward larger temporal buffers with greater overlap among mid ranks. This is consistent with weaker profile separation in this indicator space.
Rank~1 (\emph{risk taker / high urgency}) represents the smallest $\mathrm{T}_2$ and lowest CTTC values.
Ranks~2--3 (\emph{assertive to typical}) exhibit relatively close $\mathrm{T}_2$ and CTTC levels. This implies incremental differences in near-miss urgency rather than sharply separated stance categories.
Rank~4 (\emph{cautious}) exhibits larger $\mathrm{T}_2$ and CTTC, consistent with more buffered conflict timing and weaker closing urgency.
Rank~5 (\emph{very cautious / risk averse}) is distinct with much larger $\mathrm{T}_2$ and higher CTTC. This is consistent with sessions where minimum timing and closing urgency remain comparatively large.
From a safety perspective, these hypothesized stances matter because they summarize how often a deployment context produces sessions whose most critical moment occurs under tight temporal margins and urgent closing. Small shifts in average indicator values can be misleading when risk is driven by the upper tail. In contrast, profile prevalence directly quantifies whether a larger share of encounters falls into regimes consistent with risk accepting or highly buffered behaviour. This makes the results actionable for evaluation and design. If AV-only scenarios increase the prevalence of high urgency ranks in a given location, the concern is not a uniform safety degradation, but a heavier tail of time-critical encounters that can dominate safety-relevant outcomes. Conversely, higher prevalence of highly cautious ranks indicates more buffered interaction opportunities and lower exposure to near-miss urgency. Therefore, interpreting ranks as interaction stances provides a policy relevant lens for comparing AV and HDV interactions beyond single summary statistics and supports targeted follow up analyses that links high urgency ranks to specific scenario features (\textit{e.g.}, traffic flow, eHMI, and yielding behaviour).

\section{Conclusions}
\label{sec:conclusions}
This paper applies LPA to SSMs to identify and compare latent pedestrian--vehicle interaction risk profiles across vehicle types and locations. This profile based approach ensures that rare but consequential, time critical sessions are not masked by aggregate summaries which preserves heterogeneity in interaction criticality.
It also provides valuable insights for evaluating AV safety by identifying how AVs may affect pedestrian behaviour.
Using immersive VR data from Toronto and Newcastle, each 60~s session is represented by its most critical time instance, defined by the session minimum of MTTC (Model~A) or $\mathrm{T}_2$ (Model~B), paired with the corresponding CTTC at the same time step. 
Key results indicate that the Newcastle data shows clearer separation between latent risk profiles. In Toronto Model~B, the separation is less clear likely due to the indicator structure. Additionally, in Newcastle, AV-only sessions led to more high risk interactions compared to HDV-only sessions. This suggests that AVs introduce higher levels of risk in some contexts. In contrast, Toronto did not show this trend, indicating that location and context play a significant role in the impact of AVs on pedestrian risk.
The resulting profiles form an ordered risk rank that spans from high urgency (small temporal margins and urgent closing) to highly buffered encounters. These ranks likely reflect distinct pedestrian crossing stances, ranging from risk accepting to highly cautious. We can better interpret how interaction criticality varies across contexts through analyzing these profiles.
Substantively, Newcastle shows a clear shift toward higher urgency profiles in AV-only sessions. In the $\mathrm{T}_2$-based model, this shift remains informative for near-miss encounters. For Toronto, there was no distinguishable differences between AV-only and HDV-only sessions. This is likely due to weaker separation in the $\mathrm{T}_2$-based model, which captures a more continuous urgency spectrum. These findings suggest that AV-related effects are context-specific, and that conclusions depend heavily on how interaction criticality is defined.
Future work should explicitly model the influence of control variables embedded in the VR experimental design. These include traffic flow, weather, time of day, pedestrian density, social conformity, median presence, eHMI availability, and AV yielding behaviour. Including these factors into a joint modelling framework will allow assessment of how specific design and operational features shift sessions between latent risk profiles, rather than only comparing aggregate prevalence by vehicle type. Extending the framework to hybrid models that link latent profile membership to behavioural and contextual covariates would further support hypothesis driven evaluation of AV design choices and enable more direct translation of profile based findings into safety guidance for AV deployment.

\bibliographystyle{IEEEtran}
\bibliography{IEEEfull}

@inproceedings{Chen2020,
  author    = {Chen, Wenxiang and Jiang, Qianni and Zhuang, Xiangling and Ma, Guojie},
  title     = {Comparison of Pedestrians' Gap Acceptance Behavior Towards Automated and Human-Driven Vehicles},
  booktitle = {Engineering Psychology and Cognitive Ergonomics. Cognition and Design},
  series    = {Lecture Notes in Computer Science},
  volume    = {12187},
  pages     = {253--261},
  year      = {2020},
  publisher = {Springer International Publishing},
  doi       = {10.1007/978-3-030-49183-3_20}
}

@article{Velasco2021,
  author    = {Nu{\~n}ez Velasco, Juan Pablo and Lee, Yee Mun and Uttley, Jim and Solernou, Albert and Farah, Haneen and van Arem, Bart and Hagenzieker, Marjan and Merat, Natasha},
  title     = {Will pedestrians cross the road before an automated vehicle? The effect of drivers' attentiveness and presence on pedestrians' road crossing behavior},
  journal   = {Transportation Research Interdisciplinary Perspectives},
  volume    = {12},
  pages     = {100466},
  year      = {2021},
  publisher = {Elsevier},
  doi       = {10.1016/j.trip.2021.100466}
}

@techreport{LaureshynVarhelyi2018,
  author      = {Laureshyn, Aliaksei and V{\'a}rhelyi, Andr{\'a}s},
  title       = {The Swedish Traffic Conflict Technique: Observer's Manual},
  institution = {Lund University},
  address     = {Lund, Sweden},
  year        = {2018}
}

@article{Nazemi2025,
  author  = {Nazemi, Mohsen and Rababah, Bara and Ramos, Daniel and Zhao, Tangxu and Farooq, Bilal},
  title   = {Decoding pedestrian stress on urban streets using electrodermal activity monitoring in virtual immersive reality},
  journal = {Transportation Research Part C: Emerging Technologies},
  volume  = {171},
  pages   = {104952},
  year    = {2025},
  doi     = {10.1016/j.trc.2024.104952}
}

@article{Laureshyn2010,
  author  = {Laureshyn, Aliaksei and Svensson, {\AA}se and Hyd{\'e}n, Christer},
  title   = {Evaluation of traffic safety, based on micro-level behavioural data: Theoretical framework and first implementation},
  journal = {Accident Analysis \& Prevention},
  volume  = {42},
  number  = {6},
  pages   = {1637--1646},
  year    = {2010},
  doi     = {10.1016/j.aap.2010.03.021}
}

@article{Ozbay2008,
  author  = {Ozbay, K. and Yang, H. and Bartin, B. and Mudigonda, S.},
  title   = {Derivation and Validation of New Simulation-Based Surrogate Safety Measure},
  journal = {Transportation Research Record},
  volume  = {2083},
  number  = {1},
  pages   = {105--113},
  year    = {2008},
  doi     = {10.3141/2083-12}
}

@article{alhaideri2025cttc,
  title        = {Beyond traditional models: a discrete choice approach to vehicle interaction modelling at roundabouts},
  author       = {Al-Haideri, Rulla and Ismail, Karim and Weiss, Adam},
  journal      = {Transportmetrica A: Transport Science},
  pages        = {1--45},
  year         = {2025},
  doi          = {10.1080/23249935.2025.2552982}
}

@article{NaginOdgers2010,
  title        = {Group-based trajectory modeling in clinical research},
  author       = {Nagin, Daniel S. and Odgers, Candice L.},
  journal      = {Annual Review of Clinical Psychology},
  volume       = {6},
  pages        = {109--138},
  year         = {2010},
  doi          = {10.1146/annurev.clinpsy.121208.131413}
}

@article{HosseiniShoabjarehMamdoohiNordfjaern2021,
  title        = {Analysis of pedestrians' behaviour: A segmentation approach based on latent variables},
  author       = {Hosseini Shoabjareh, A. and Mamdoohi, A. R. and Nordfj{\ae}rn, T.},
  journal      = {Accident Analysis \& Prevention},
  volume       = {157},
  pages        = {106160},
  year         = {2021},
  doi          = {10.1016/j.aap.2021.106160}
}

@article{MajiGhosh2025,
  title        = {A copula-based multivariate extreme value framework for roundabout safety evaluation under mixed traffic},
  author       = {Maji, A. and Ghosh, I.},
  journal      = {Accident Analysis \& Prevention},
  volume       = {221},
  pages        = {108219},
  year         = {2025},
  doi          = {10.1016/j.aap.2025.108219}
}

@article{vanLissa2018tidyLPA,
  author  = {Rosenberg, Joshua M. and Beymer, Patrick N. and Anderson, Daniel J. and van Lissa, Caspar J. and Schmidt, Jennifer A.},
  title   = {{tidyLPA}: An {R} Package to Easily Carry Out Latent Profile Analysis ({LPA}) Using Open-Source or Commercial Software},
  journal = {Journal of Open Source Software},
  volume  = {3},
  number  = {30},
  pages   = {978},
  year    = {2018},
  doi     = {10.21105/joss.00978}
}

\end{document}